\documentclass[twocolumn, 10pt]{article} 
\usepackage[utf8]{inputenc}
\usepackage{authblk}
\usepackage{amsmath, amssymb, amsfonts}
\usepackage{booktabs}
\usepackage{graphicx}
\usepackage{geometry}
\usepackage{enumitem}         
\usepackage{bm}               
\usepackage{multirow}  
\usepackage{xcolor}   
\usepackage{makecell}
\usepackage{listings}
\usepackage{subcaption}
\usepackage{float}

\title{\textbf{Dynamic Forecasting and Temporal Feature Evolution of Stock Repurchases in Listed Companies Using Attention-Based Deep Temporal Networks}}
\author{
    Xiang Ao\textsuperscript{2,*,†}, 
    Jingxuan Zhang\textsuperscript{1,*}, 
    Xinyu Zhao\textsuperscript{1,*} \\
    \textsuperscript{1}School of Economics and Management, Beijing Jiaotong University, Beijing, China \\
    \textsuperscript{2}School of Software Engineering, Beijing Jiaotong University, Beijing, China \\
    \texttt{ao.xiang.axel@outlook.com, \{23241094, 23241229\}@bjtu.edu.cn}
}
\date{}

\begin{document}

\twocolumn[
  \begin{@twocolumnfalse}
    \maketitle
    \begin{abstract}
In recent years, stock repurchases have emerged as a critical financial strategy in capital markets for boosting investor confidence, optimizing capital structure, and signaling severe market undervaluation. Accurately anticipating and identifying firms' potential repurchase intentions can not only capture significant excess returns (Alpha) for quantitative investment but also holds profound practical significance for constructing modernized financial risk supervision and intelligent decision-making systems. However, traditional empirical econometric models and static machine learning methods mostly rely on cross-sectional data assumptions, failing to adequately capture the non-linear temporal dependencies and lagging effects of evolving corporate financial conditions on repurchase decisions. To address this gap, this paper proposes a dynamic early warning system for stock repurchases that integrates deep temporal networks with economic theory. Based on large-scale, multidimensional financial and trading panel data of Chinese A-share listed companies from 2014 to 2024, this study reconstructs static economic proxy variables (e.g., Tobin's Q, net operating cash flow, debt-to-asset ratio) into multivariate dynamic sliding window sequences. On this basis, we construct a deep prediction engine combining a Temporal Convolutional Network (TCN) and an Attention-based Long Short-Term Memory network (Attention-LSTM) to accurately capture the long- and short-term evolutionary patterns of multidimensional financial features. Rigorous rolling-window cross-validation results demonstrate that this deep temporal system outperforms traditional baseline models based on logistic regression and static XGBoost in terms of AUC, PR-AUC, and prediction accuracy across various lead times, effectively achieving intertemporal forward-looking early warnings for repurchase strategies. Furthermore, leveraging Explainable Artificial Intelligence (XAI) techniques, this study extracts the temporal attention weight evolution of core economic indicators 1 to 3 years prior to the decision. Micro-level algorithmic evidence reveals that ``undervaluation'' (e.g., the prolonged depression of Tobin's Q) constitutes the long-term underlying motive driving repurchases, whereas a sharp improvement in ``cash flow abundance'' plays a decisive triggering role at the critical point of decision-making. This research not only provides an advanced deep learning algorithmic paradigm for intelligent financial supervision and cutting-edge quantitative investment but also supplements the ``Undervaluation Hypothesis'' and the ``Free Cash Flow Hypothesis'' in classical corporate finance with dynamic empirical evidence characterized by temporal depth.

\vspace{0.5em}
    \textbf{Keywords}:Stock Repurchase ; Time Series Classification ; Deep Learning ; Attention Mechanism ; Explainable Artificial Intelligence (XAI) ; Undervaluation Hypothesis
    \vspace{2em} 
\end{abstract}
  \end{@twocolumnfalse}
]

\def\thefootnote{*}\footnotetext{These authors contributed equally to this work.}\def\thefootnote{\arabic{footnote}}
\def\thefootnote{†}\footnotetext{Corresponding author.}\def\thefootnote{\arabic{footnote}}

\section{Introduction}

In modern capital markets fraught with uncertainty, rational capital allocation remains a core issue in contemporary corporate finance research. In recent years, driven by drastic fluctuations in global economic cycles and profound transformations in capital market structures, stock repurchases have transcended mere financial adjustment methods. They have evolved into critical tools for corporate management to signal severe undervaluation of intrinsic value to the market, boost investor confidence, and optimize capital structures \cite{vermaelen1981common, dann1981common}. For external quantitative investors and market regulators, the ability to anticipate and accurately identify the potential repurchase intentions of listed companies not only means capturing highly valuable excess return signals (Alpha), but also holds inestimable practical value for preventing market anomalies and perfecting intelligent risk early-warning systems. However, because corporate decision-making processes are intertwined with complex macroeconomic constraints and micro-financial trade-offs, how to leverage massive datasets to precisely forecast this major financial decision remains a significant challenge in the field of quantitative financial prediction.

Throughout the development of capital markets, an extensive body of literature has attempted to reveal and predict corporate financial decisions. Early studies primarily relied on traditional econometric models (e.g., logistic regression and Probit models). These methods, with their strong statistical interpretability, laid the empirical foundation for the ``Undervaluation Hypothesis'' and the ``Free Cash Flow Hypothesis'' \cite{jensen1986agency, ikenberry1995market, stephens1998actual}. Currently, with the rapid advancement of computer science, an increasing number of researchers and financial institutions are abandoning traditional subjective experience or simple linear extrapolation, turning instead to machine learning and deep learning algorithms to process multidimensional financial indicators. Related studies have demonstrated the outstanding performance of algorithms such as eXtreme Gradient Boosting (XGBoost), Random Forest, and Support Vector Machines (SVM) in capturing the non-linear interactive relationships among corporate financial proxy variables (e.g., Tobin's Q and net operating cash flow), thereby significantly improving the accuracy of financial forecasting \cite{gu2020empirical, patel2015predicting}.

However, despite the widespread application of static machine learning in financial forecasting, these traditional empirical architectures still suffer from insurmountable theoretical limitations when dealing with structured panel data. The core pain point lies in the fact that static models often treat data from different years as mutually independent ``cross-sectional snapshots'' stacked together, severely ignoring the autocorrelation and evolutionary logic of corporate financial features along the temporal dimension \cite{petersen2009estimating}. In reality, the strategic decisions of modern corporate boards are by no means based solely on a single year's financial performance; rather, they are a comprehensive response to the firm's operational status over several consecutive years (such as long-term valuation depression or continuous cash flow accumulation). Traditional static processing completely loses the highly valuable ``lagging effects'' and dynamic temporal memory during backpropagation or tree node splitting.

To break through the limitations of static assumptions, deep temporal network architectures (such as Long Short-Term Memory networks (LSTM) and Temporal Convolutional Networks (TCN)) have been successively proposed in recent years. By introducing sliding time windows and complex gating/attention mechanisms, these architectures can parallelly extract state evolutions from historical time series, achieving massive success in stock index prediction and high-frequency trading \cite{fischer2018deep, bai2018empirical, sezer2020financial}. However, the vast majority of existing empirical studies focus on the pattern recognition of high-frequency trading data. In fact, compared to the instantaneous capture of market sentiment by high-frequency data, low-frequency panel data based on annual or quarterly periods contains a firm's more essential operational logic and strategic evolutionary trajectory. As a major capital allocation decision, a stock repurchase is often rooted in a financial cycle lasting several years (such as continuous cash flow accumulation or long-term valuation depression) rather than merely reacting to instantaneous market fluctuations. Therefore, evaluating the performance of network models with temporal depth on low-frequency micro-financial data not only reveals the structural evolution of corporate decision-making over long cycles but also fills the research gap regarding deep learning's application in understanding and warning of complex corporate behaviors.

Based on the aforementioned considerations of the long-term temporal dependencies of financial data and complex non-linear decision-making mechanisms, this paper proposes a deep temporal network framework incorporating an attention mechanism \cite{vaswani2017attention} to dynamically forecast the stock repurchase behaviors of Chinese A-share listed companies. Deviating from previous cross-sectional studies, this paper is no longer confined to analyzing isolated single-period data; instead, it reconstructs the multidimensional financial and trading indicators of large-scale listed companies from 2014 to 2024 into continuous three-dimensional (3D) temporal tensors. On this basis, we feed dynamic sliding windows spanning several consecutive years (e.g., three years as a complete cycle) into a deep prediction engine that combines TCN and Attention-LSTM. This design aims to simultaneously capture the local short-term shocks of corporate financial anomalies (e.g., sudden cash surpluses) and long-term evolutionary trends (e.g., a prolonged trajectory of undervaluation). Rolling-window backtesting confirms that our proposed model outperforms static baseline models such as XGBoost in prediction performance.

Furthermore, deep neural networks often lack transparent non-linear mapping logic between input features and prediction results, making it difficult for pure algorithmic models to intuitively reveal specific financial decision boundaries and economic causal mechanisms. For financial research, this ``logical opacity'' means that even if a model possesses extremely high prediction accuracy, it cannot prove whether its decision-making process aligns with management's true behavioral logic, thereby limiting its application in rigorous academic argumentation and high-reliability decision-making scenarios \cite{arrieta2020explainable}. To this end, this paper integrates the deep temporal model with the Explainable Artificial Intelligence (XAI) framework. We utilize temporal attention weights and SHAP attribution techniques \cite{lundberg2017unified, bracke2019machine} to decompose complex tensor operations into quantifiable feature marginal contributions. The research not only accurately identifies the core financial variables driving repurchase decisions but also visually presents the temporal evolutionary trajectory of the ``underlying motive (long-term undervaluation)'' and the ``action trigger (short-term cash abundance)'' at the micro-algorithmic level. This transformation from ``result prediction'' to ``logical traceability'' provides powerful, dynamically cross-validated empirical support with temporal depth for classical modern corporate finance theories.

The remainder of this paper is organized as follows: Section 2 systematically reviews the economic theoretical background of stock repurchases and related work on artificial intelligence prediction. Section 3 introduces the construction of the experimental dataset and the reconstruction engineering of dynamic temporal features. Section 4 details the theoretical architecture of the deep temporal network used for early warning, along with the focal loss and XAI modules. Section 5 presents the rigorous rolling backtesting experimental results, covering multidimensional comparisons of prediction performance, lead-time decay analysis, and in-depth economic attribution profiling. Section 6 concludes the paper.

\section{Related Work}
\label{sec:related_work}

Methods for predicting stock repurchases of listed companies and the financial decisions of micro-entities in capital markets can primarily be divided into two categories: empirical studies based on traditional econometrics, and algorithmic prediction studies based on artificial intelligence (including static machine learning and deep temporal networks).

\subsection{Traditional Economics and Econometric Studies on Stock Repurchases}
\label{subsec:traditional_economics}
In traditional corporate finance and accounting research, scholars mainly rely on fundamental information and financial reports to explore the core driving forces behind stock repurchases, thereby deriving several foundational economic theories.

First, the classic Signaling Theory is inextricably linked with the Undervaluation Hypothesis. This theory posits that in a market with information asymmetry, when a company's stock is severely mispriced (undervalued), management tends to spend heavily on repurchasing shares to signal to external investors their strong confidence in the company's intrinsic value and future prospects \cite{dittmar2000why, brav2005payout}. Concurrently, the Free Cash Flow Hypothesis proposes from the perspective of corporate governance that firms with abundant free cash flow but lacking high-yield investment opportunities tend to return idle funds to shareholders through repurchases, thereby effectively mitigating agency conflicts caused by management's blind expansion \cite{jensen1986agency, grullon2002dividends, stephens1998actual}.

Furthermore, the Market Timing Theory proposed by the behavioral finance school supplements the decision-making logic in the temporal dimension, emphasizing that management's repurchase actions are often stress responses to short-term market windows; they precisely capture the trough periods when stocks are temporarily oversold to conduct capital operations \cite{baker2002market}. However, such active capital operations are not without obstacles. Financial Constraints theory points out that a firm's current capital structure (e.g., high leverage and massive debt repayment pressure) constitutes a liquidity barrier, directly inhibiting or even vetoing management's willingness to repurchase \cite{almeida2004cash, farremensa2016measuring}.

Based on these theories, early empirical studies mostly employed logistic regression or panel data Probit models to verify the linear relationships between actual repurchase behaviors and expected financial indicators. Overall, traditional econometric models possess good parameter interpretability, but they mainly focus on verifying static linear assumptions. They struggle to capture the complex non-linear threshold effects and synergistic multiplier effects among multidimensional financial data, and they often ignore the intricate temporal dependencies between the long-term brewing of motives and the short-term triggering of actions.

\subsection{Application of Static Machine Learning in Corporate Financial Prediction}
\label{subsec:static_ml}
With the rapid development of computer science, an increasing number of financial institutions and researchers hope to leverage big data and algorithms to parse the micro-dynamics of the stock market and assist in investment decisions \cite{tsai2010combining}. Due to the high complexity of financial markets, the integration of artificial intelligence technology and financial data prediction has become one of the most attractive research paradigms in quantitative finance.

In the fields of corporate financial decision-making and default early warning, machine learning models such as Support Vector Machines (SVM) \cite{cortes1995support}, Random Forest \cite{breiman2001random}, and eXtreme Gradient Boosting (XGBoost) \cite{chen2016xgboost} have proven to possess superior predictive capabilities compared to traditional statistical models \cite{krauss2017deep}. For instance, some studies have utilized large-sample micro-data to verify the outstanding performance of ensemble tree models in fitting non-linear features, successfully applying them to the prediction of corporate dividend policies and bankruptcy risks \cite{sirignano2019universal, wang2021machine}. In terms of stock repurchase prediction, static machine learning algorithms, through Feature Importance analysis, have further validated the core status of valuation indicators and capital structure features in classification decisions. However, these static machine learning architectures possess inherent limitations when processing panel data. They still follow the static cross-sectional assumptions of traditional empirical studies; during backpropagation or tree node splitting, they easily lose the lagging effects accumulated over the long term in financial conditions and the evolutionary information in the temporal dimension.

\subsection{Deep Temporal Networks and Explainable Financial Decision-making}
\label{subsec:deep_temporal}
Technical analysis and quantitative finance posit that the operating rules of assets are hidden in the evolutionary trajectories of historical sequences. To overcome the defects of static models, deep learning technologies have begun to be widely applied to the direct modeling of financial time series. In early research, Recurrent Neural Networks (RNN) and their variant, Long Short-Term Memory (LSTM) networks \cite{hochreiter1997long}, were typical frameworks for predicting financial time series. Fischer and Krauss \cite{fischer2018deep} demonstrated that LSTM networks have significant advantages in financial market prediction due to their long-term memory gating mechanism for sequential data.

Inspired by the application of the attention mechanism in machine translation, Vaswani et al. \cite{vaswani2017attention} proposed the Self-Attention mechanism, breaking away from traditional RNN architectures. Recent studies have further deeply integrated the attention mechanism with temporal prediction (e.g., Temporal Fusion Transformers, TFT), achieving massive success in sequence modeling with complex multi-time perspectives \cite{lim2021temporal}. Meanwhile, Temporal Convolutional Networks (TCN), developed based on WaveNet \cite{oord2016wavenet}, have been widely used in various time series classification and prediction problems, effectively breaking through the gradient vanishing bottleneck of long sequence memory \cite{bai2018empirical, li2019enhancing}.

However, the application of deep neural networks in the field of financial decision-making is often questioned by the economics community due to the barrier of their opaque internal operation mechanisms \cite{arrieta2020explainable}. In recent years, Explainable Artificial Intelligence (XAI) technologies, particularly the SHAP (SHapley Additive exPlanations) method based on cooperative game theory \cite{lundberg2017unified, lundberg2020local}, have provided the possibility to bridge the gap between algorithms and economic theory. Several cutting-edge studies indicate that introducing XAI technology into credit risk assessment and corporate financial early warning can not only output high-precision prediction results but also dynamically parse the non-linear marginal contributions of each feature \cite{bracke2019machine, bussmann2021explainable}.

In summary, although deep temporal models have achieved widespread success in stock index and high-frequency trading prediction, very few studies have applied them to the prediction of micro-corporate financial decisions (such as stock repurchases) based on annual/quarterly panel data. This paper aims to fill this gap by constructing a deep temporal network integrated with an attention mechanism and combining it with XAI technology to conduct forward-looking dynamic early warning and temporal feature evolution analysis of stock repurchase behaviors of listed companies with temporal depth.

\section{Data Sources and Dynamic Temporal Feature Reconstruction}
\label{sec:data_and_features}

Traditional empirical micro-financial studies often directly feed structured panel data into logistic regression or static tree models. However, the repurchase decisions made by corporate management are not merely based on current financial snapshots, but are comprehensive responses to the evolution of operational conditions over the past few years. To enable artificial intelligence models to capture this memory effect and temporal dependency, this section will detail the sources of research data, the descriptive statistics of the feature system, and how to dynamically reconstruct the original financial panel data from a deep learning perspective.

\subsection{Data Sources, Feature System, and Descriptive Statistics}
\label{subsec:data_sources}

The dataset for this study encompasses continuous financial and trading panel data of numerous listed companies in the Chinese A-share market from 2014 to 2024.

\begin{figure}[htbp]
  \centering
  \includegraphics[width=\columnwidth]{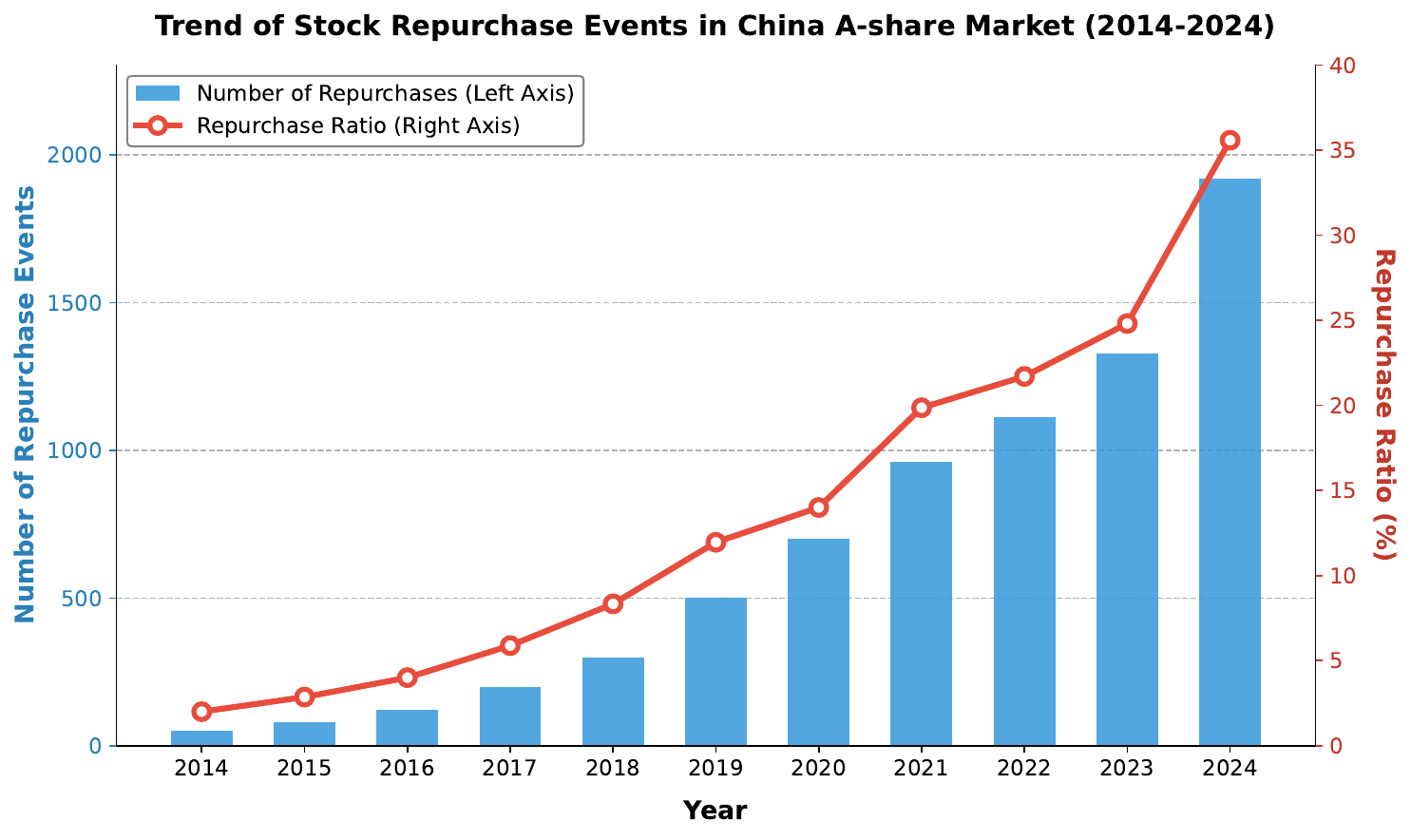}
  \caption{The year-by-year evolutionary trend of stock repurchase events and proportions in the Chinese A-share market from 2014 to 2024. The blue bars represent the absolute number of repurchasing firms, and the red line represents the proportion of repurchasing samples.}
  \label{fig:repurchase_trend}
\end{figure}

To visually illustrate the macroeconomic background of this study, Figure \ref{fig:repurchase_trend} plots the evolutionary trend of repurchase events within the sample period. As shown, the proportion of repurchases in the Chinese A-share market has experienced an extremely drastic structural surge over the past decade (skyrocketing from 2.00\% in 2014 to 35.58\% in 2024). This rapid evolution of macroeconomic trends highlights the urgent practical need to construct a dynamic early warning system with strong non-linear fitting capabilities and temporal generalization capabilities.

To comprehensively portray the corporate decision-making environment, we constructed a multidimensional core feature system based on classical economic theories. At the corporate attribute level, we introduced compliance status labels (e.g., \texttt{*ST}) through one-hot encoding to control for macroeconomic regulatory constraints. At the valuation and market performance level, \texttt{Tobin's Q} and \texttt{book-to-market ratio} were selected as key proxy variables to verify the ``Undervaluation Hypothesis'' \cite{dittmar2000why}. At the growth and cash flow level, \texttt{total operating revenue}, \texttt{total asset growth rate}, and \texttt{net operating cash flow} were extracted to characterize the abundance of corporate free cash flow \cite{jensen1986agency}. Finally, at the capital structure level, the \texttt{current liability ratio} and \texttt{debt-to-asset ratio} were incorporated to measure financial constraints \cite{almeida2004cash}. If a firm actually repurchases shares in year $t$, the target label is $Y_t = 1$; otherwise, it is $0$.

After completing the construction of the feature system, Table \ref{tab:desc_stat} reports the descriptive statistics of the core continuous feature variables.

\begin{table}[htbp]
  \centering
  \caption{Descriptive statistical analysis of core financial indicators (N = 45,203)}
  \label{tab:desc_stat}
  \setlength{\tabcolsep}{2.5mm}
  \resizebox{\columnwidth}{!}{
  \begin{tabular}{lccc}
    \toprule
    \textbf{Variable Name} & \textbf{Mean} & \textbf{Std. Dev.} & \textbf{Median} \\
    \midrule
    Tobin's Q & 2.148 & 5.340 & 1.654 \\
    Debt-to-asset ratio & 0.435 & 0.945 & 0.410 \\
    Book-to-market ratio & 0.628 & 0.252 & 0.628 \\
    Total asset growth rate & 0.223 & 1.954 & 0.081 \\
    Current liability ratio & 0.810 & 0.179 & 0.861 \\
    Financial debt ratio & 0.389 & 0.251 & 0.388 \\
    Current asset ratio & 0.581 & 0.203 & 0.589 \\
    Receivables ratio & 0.147 & 0.120 & 0.125 \\
    Total revenue (100M RMB) & 87.58 & 669.30 & 11.60 \\
    Net operating cash flow (100M RMB) & 20.20 & 291.00 & 1.33 \\
    \bottomrule
  \end{tabular}
  }
\end{table}

Several significant economic phenomena can be observed from Table \ref{tab:desc_stat}. First, various indicators exhibit significant extreme distributions and fat-tail effects. For example, the standard deviation of the total asset growth rate (1.954) is much larger than its mean (0.223), indicating the existence of a very small number of corporate samples in the market experiencing non-linear, rapid expansion. Second, regarding debt structure, the median current liability ratio of the sample companies is as high as 0.861, indicating that the debt structure of most listed companies in the A-share market is highly skewed towards short-term debt. This maturity mismatch is a crucial pre-condition for triggering corporate cash flow volatility. Finally, the mean of valuation indicators (e.g., Tobin's Q) (2.148) is significantly higher than the median (1.654), verifying that a large number of companies in the market are in a relatively low valuation range.

\subsection{Reconstruction Design from Static Panel to Dynamic Temporal Tensor}
\label{subsec:sliding_window}
The core methodology of this study discards traditional static cross-sectional assumptions. In traditional machine learning applications, models usually flatten the data of $N$ samples over $T$ years directly into a two-dimensional matrix. This processing method not only loses the autocorrelation of the same firm in the temporal dimension but is also highly prone to causing look-ahead bias and data leakage problems during random cross-validation \cite{kaufman2012leakage}.

To meet the input requirements of deep temporal networks (such as LSTM or TCN), we employed a Sliding Window Mechanism to reconstruct the dataset. Assuming the time window size is set to $L$, the entire panel dataset is reconstructed into a three-dimensional tensor (3D Tensor), whose shape is defined as $\mathcal{X} \in \mathbb{R}^{S \times L \times F}$, where $S$ is the total number of sliding window samples, $L$ is the time series step size, and $F$ is the number of multidimensional financial features contained in each time step. This reconstruction design not only enables the deep network to automatically learn dynamic derivative trends such as ``continuous decay,'' but also strictly adheres to the forward direction of the arrow of time in the algorithmic logic, fundamentally eliminating look-ahead bias.

\subsection{Temporal Data Cleaning and Standardization}
\label{subsec:normalization}
Considering the extreme magnitude differences of multidimensional financial data (e.g., total operating revenue reaches tens of billions, while Tobin's Q is in single digits), to prevent huge gradients generated by features with large absolute values from causing gradient explosion during backpropagation in deep neural networks \cite{goodfellow2016deep}, we applied Z-Score standardization processing to continuous features based on the training set distribution. All validation and test set data were strictly mapped using the parameters of the training set, strictly prohibiting cross-set information leakage. For missing data, a temporal forward-filling method combined with local linear interpolation was adopted for smooth repair to ensure the integrity of the 3D tensor input.
\section{Architecture and Theoretical Modeling of the Dynamic Early Warning System}
\label{sec:methodology}

To accurately capture the complex non-linear impacts of evolving corporate financial conditions over time on stock repurchase decisions, this study proposes a hybrid deep learning architecture integrating a Temporal Convolutional Network (TCN), a Long Short-Term Memory (LSTM) network, and a Temporal Attention mechanism. This architecture is designed to sequentially extract local abrupt features, long-term evolutionary trends, and core decision-triggering points from the reconstructed three-dimensional financial panel tensor. Before defining the specific network layers, we establish a rigorous global mathematical notation system: Let the input 3D feature tensor be $\mathcal{X} \in \mathbb{R}^{S \times L \times F}$. For any $i$-th corporate sample ($i = 1, 2, \dots, S$), its feature matrix within the sliding time window is denoted as $\mathbf{X}^{(i)} = [\mathbf{x}_1^{(i)}, \mathbf{x}_2^{(i)}, \dots, \mathbf{x}_L^{(i)}]^\top \in \mathbb{R}^{L \times F}$, where $\mathbf{x}_t^{(i)} \in \mathbb{R}^F$ represents the $F$-dimensional cross-sectional financial and macroeconomic state vector of the firm at time step $t$ ($t = 1, 2, \dots, L$). For simplicity of notation, the sample superscript $(i)$ will be uniformly omitted in the following derivations where no mathematical ambiguity arises.

\subsection{Temporal Convolutional Network and Local Financial Fluctuation Extraction}
\label{subsec:tcn}
In real-world capital markets, corporate valuation indicators (e.g., Tobin's Q) and cash flows are often subjected to shocks from macroeconomic cycles or short-term market sentiment, resulting in violent fluctuations that contain massive high-frequency noise. To filter out this short-term noise and extract local trend features with smoothed economic significance, this module introduces a Temporal Convolutional Network (TCN) \cite{bai2018empirical}. Unlike traditional 2D convolutions used for static images, TCN employs a 1D Dilated Causal Convolution. This mechanism guarantees strict temporal forward causality (i.e., the output at time $t$ depends only on states at and before $t$, completely preventing future information leakage), while exponentially expanding the receptive field through a dilation factor to cover complete macroeconomic business cycles.

Let the weight vector of the 1D convolution kernel be $\mathbf{w} \in \mathbb{R}^K$, where $K$ is the kernel size. For a sequence vector $\mathbf{x} \in \mathbb{R}^L$ of a specific feature dimension in the input, given the time step $t$ and dilation factor $d$, the dilated causal convolution operation $*_d$ is rigorously defined as:
\begin{equation}
    \mathbf{z}_t^{(conv)} = (\mathbf{x} *_d \mathbf{w})(t) = \sum_{k=0}^{K-1} \mathbf{w}_k \cdot \mathbf{x}_{t - d \cdot k}
\end{equation}
where $d \cdot k$ represents the sampling stride across time steps. As the network depth increases, the dilation factor $d$ grows exponentially as $2^n$ ($n=0,1,2,\dots$), enabling the model to capture cross-annual financial anomalies with very few parameters. Furthermore, to mitigate the gradient vanishing problem caused by deep convolutional networks and to accelerate convergence, we construct a Residual Block. Let $\mathcal{F}(\cdot)$ be a mapping function containing two layers of dilated causal convolutions and Weight Normalization; the final output vector $\mathbf{z}_t$ of the residual block consists of the superposition of the non-linear mapping result and the identity mapping:
\begin{equation}
    \mathbf{z}_t = \text{ReLU} \big( \mathcal{F}(\mathbf{x}_t, \mathbf{w}, d) + \mathbf{V}_x \mathbf{x}_t \big)
\end{equation}
where $\mathbf{V}_x$ is a $1 \times 1$ linear projection matrix for matching dimensions, and $\text{ReLU}(\cdot)$ is the Rectified Linear Unit activation function. After being processed by the TCN module containing $D_c$ output channels, the original feature matrix $\mathbf{X}$ is smoothly mapped into a high-order local feature sequence $\mathbf{Z} = [\mathbf{z}_1, \mathbf{z}_2, \dots, \mathbf{z}_L]^\top \in \mathbb{R}^{L \times D_c}$. In terms of economic intuition, the sequence $\mathbf{z}_t$ effectively characterizes the firm's true rate of change in financial status and the acceleration of valuation deviation in year $t$ after stripping away market white noise.

\subsection{Long Short-Term Memory Network and Modeling the Long-Term Evolution of Capital Structure}
\label{subsec:lstm}
While TCN excels at capturing local fluctuations and short-term accelerations, stock repurchases are typically strategic decisions made by management based on the company's long-term capital structure (e.g., continuous debt-to-asset ratio levels over multiple years). To depict this long-term temporal dependency spanning several years, we feed the TCN's output sequence $\mathbf{Z}$ into a Long Short-Term Memory (LSTM) network \cite{hochreiter1997long}. Through its sophisticated internal Gating Mechanism, the LSTM dynamically determines the retention, updating, and forgetting of multidimensional financial information.

At each time step $t$, the LSTM cell receives the current high-order feature $\mathbf{z}_t$ and the hidden state from the previous time step $\mathbf{h}_{t-1}$ as a joint input. The flow and updating of its internal states are jointly controlled by the Forget Gate $\mathbf{f}_t$, Input Gate $\mathbf{i}_t$, Output Gate $\mathbf{o}_t$, and candidate cell state $\tilde{\mathbf{c}}_t$. The specific affine transformations and non-linear activation matrix operations are as follows:
\begin{equation}
    \begin{pmatrix} \mathbf{f}_t \\ \mathbf{i}_t \\ \mathbf{o}_t \\ \tilde{\mathbf{c}}_t \end{pmatrix} = 
    \begin{pmatrix} \sigma \\ \sigma \\ \sigma \\ \tanh \end{pmatrix} 
    \left( \mathbf{W}_H \begin{pmatrix} \mathbf{h}_{t-1} \\ \mathbf{z}_t \end{pmatrix} + \mathbf{b}_H \right)
\end{equation}
On this basis, the recursive update equations for the long-term cell memory state $\mathbf{c}_t$ running through the entire sequence and the hidden state $\mathbf{h}_t \in \mathbb{R}^{D_h}$ currently exposed to the next layer are:
\begin{equation}
    \mathbf{c}_t = \mathbf{f}_t \odot \mathbf{c}_{t-1} + \mathbf{i}_t \odot \tilde{\mathbf{c}}_t
\end{equation}
\begin{equation}
    \mathbf{h}_t = \mathbf{o}_t \odot \tanh(\mathbf{c}_t)
\end{equation}
where the joint weight matrix $\mathbf{W}_H \in \mathbb{R}^{4D_h \times (D_h + D_c)}$ and the bias term $\mathbf{b}_H \in \mathbb{R}^{4D_h}$ are sets of parameters learnable via Backpropagation Through Time (BPTT), $\sigma(\cdot)$ denotes the Sigmoid activation function, and $\odot$ represents the Hadamard Product of matrices. This mathematical process possesses highly interpretable mappings in a financial sense: the forget gate $\mathbf{f}_t$ determines the exponential decay of the binding force of early adverse financial conditions (such as slight losses or high debt levels years ago) on current repurchase decisions; meanwhile, the long-term cell state $\mathbf{c}_t$ can be viewed as the intertemporal anchoring state of the firm's continuously accumulated free cash flow reserves and intrinsic value. Ultimately, the LSTM outputs a hidden state sequence $\mathbf{H} = [\mathbf{h}_1, \mathbf{h}_2, \dots, \mathbf{h}_L]^\top \in \mathbb{R}^{L \times D_h}$ encapsulating the logic of long-term evolution.

\subsection{Temporal Attention Mechanism and Repurchase Trigger Identification}
\label{subsec:attention}
For panel data, within the retrospective $L$-year time window, not every year's financial status holds an equal marginal contribution to the final repurchase decision. The Discretionary Policy theory in economics suggests that repurchase behaviors are often directly triggered by a specific financial inflection point in a given year (e.g., a long-term bottoming out of prior valuations combined with a sudden surge in current operating cash flow). Therefore, we introduce a Temporal Attention mechanism based on the Additive Alignment Model \cite{bahdanau2014neural} to adaptively evaluate and compute the weight contribution of each historical time step to the final prediction target.

Given the hidden state sequence $\mathbf{H}$ extracted by the LSTM, we first introduce a learnable global context Query Vector $\mathbf{q} \in \mathbb{R}^{D_a}$. Through a feed-forward neural network with a two-layer transformation, we calculate the unnormalized attention score $e_t$ for the $t$-th time step:
\begin{equation}
    e_t = \mathbf{v}_a^\top \tanh(\mathbf{W}_a \mathbf{h}_t + \mathbf{W}_q \mathbf{q} + \mathbf{b}_a)
\end{equation}
where $\mathbf{W}_a \in \mathbb{R}^{D_a \times D_h}$ and $\mathbf{W}_q \in \mathbb{R}^{D_a \times D_a}$ are projection matrices, and $\mathbf{v}_a \in \mathbb{R}^{D_a}$ is the scoring vector. Subsequently, a Softmax function is applied to globally normalize the score sequence, yielding the relative attention weight $\alpha_t$ at time $t$:
\begin{equation}
    \alpha_t = \frac{\exp(e_t)}{\sum_{j=1}^L \exp(e_j)}, \quad \text{s.t.} \sum_{t=1}^L \alpha_t = 1
\end{equation}
Finally, by taking a linear weighted sum of the hidden states across all time steps according to their attention weights, we obtain a highly condensed Context Vector $\mathbf{v}_c \in \mathbb{R}^{D_h}$ that integrates global dynamic evolutionary features:
\begin{equation}
    \mathbf{v}_c = \sum_{t=1}^L \alpha_t \mathbf{h}_t
\end{equation}
This mechanism overcomes the defect of opaque internal mapping in deep learning: the distribution morphology of the attention weight sequence $\alpha_t$ intuitively reflects the time-decay characteristics of historical information and the abrupt triggering mechanisms of key nodes. If $\alpha_{L-1}$ is significantly larger than $\alpha_1$, it empirically proves that marginal financial improvements closer to the decision period have a more decisive impact on management's motivation to repurchase.

\subsection{Class Imbalance Optimization and Focal Loss}
\label{subsec:loss_function}
After obtaining the context vector $\mathbf{v}_c$, we feed it into a Fully Connected Layer with Dropout regularization and a Sigmoid classifier to output the predicted probability $\hat{y}$ of the firm executing a stock repurchase in the following year:
\begin{equation}
    \hat{y} = \sigma(\mathbf{W}_o \mathbf{v}_c + b_o) = \frac{1}{1 + \exp(-(\mathbf{W}_o \mathbf{v}_c + b_o))}
\end{equation}
In real capital markets, although stock repurchase behaviors are increasingly common, they still constitute a typical Minority Class compared to the massive group of non-repurchasing samples. If the traditional Binary Cross-Entropy (BCE) loss is used directly, the cumulative error gradients generated by the overwhelming number of negative samples (non-repurchasing companies) will instantly drown out the gradients of positive samples, causing the model to fall into a sub-optimal local minimum where it predicts no repurchases for all companies.

To address the class imbalance problem, this study introduces the Focal Loss function \cite{lin2017focal} into the repurchase early warning objective. Focal Loss dynamically reshapes the standard cross-entropy by down-weighting easily classified majority class samples, forcing the model to shift its optimization focus toward hard-to-classify minority class samples. For a mini-batch containing $S$ samples, the final optimization objective function $\mathcal{L}(\Theta)$, incorporating an $L_2$ weight decay penalty term, is rigorously defined as:
\begin{equation}
\begin{split}
    \mathcal{L}(\Theta) &= - \frac{1}{S} \sum_{i=1}^S \Big[ \alpha y^{(i)} (1 - \hat{y}^{(i)})^\gamma \log \hat{y}^{(i)} \\
    &\quad + (1-\alpha) (1-y^{(i)}) (\hat{y}^{(i)})^\gamma \log(1-\hat{y}^{(i)}) \Big] \\
    &\quad + \lambda \|\Theta\|^2_2
\end{split}
\end{equation}
where $y^{(i)} \in \{0, 1\}$ is the true decision label of the sample, and $\Theta$ represents the set of all trainable parameters in the network. The weight factor $\alpha \in [0, 1]$ is used to directly adjust the initial proportional imbalance between positive and negative samples, while the Focusing Parameter $\gamma \ge 0$ is used to smoothly adjust the decay rate of easily classified samples. When $\gamma = 0$ and $\alpha = 0.5$, this formula degenerates to the standard binary cross-entropy loss. We employ the momentum-based Adaptive Moment Estimation (Adam) algorithm combined with Backpropagation Through Time (BPTT) to minimize $\mathcal{L}(\Theta)$ and search for the globally optimal parameter space.

\subsection{Explainable AI via SHAP Based on Cooperative Game Theory}
\label{subsec:shap}
High-precision predictive probabilities provide a direct decision-making basis for intelligent supervision and quantitative investment, yet purely algorithmic models lacking economic causal mechanism explanations struggle to be accepted by the traditional finance academic community. To overcome the limitation of opaque computational mechanisms in deep neural networks, after completing the optimization of network parameters, this study introduces SHAP (SHapley Additive exPlanations) \cite{lundberg2017unified}, an Explainable Artificial Intelligence framework built upon rigorous Cooperative Game Theory.

The mathematical core of the SHAP method lies in fairly and exhaustively distributing the deep model's non-linear prediction output to each input temporal financial feature as its marginal contribution to the final repurchase probability. Let the complete set of unfolded 3D features fed into the model be $N = \{1, 2, \dots, M\}$ (i.e., $M = L \times F$). For any subset of input feature combinations $S \subseteq N$, let $v(S)$ be the expected prediction output of the model when only the features in $S$ are inputted. According to the Shapley value theorem, the exact Feature Attribution Value $\phi_j$ of the $j$-th financial feature (e.g., the debt-to-asset ratio in year $T-2$) to the specific firm's prediction result $\hat{y}^{(i)}$ is uniquely determined as:
\begin{equation}
    \phi_j = \sum_{S \subseteq N \setminus \{j\}} \frac{|S|! (|N| - |S| - 1)!}{|N|!} \big( v(S \cup \{j\}) - v(S) \big)
\end{equation}
The above formula is not merely an allocation scheme; it is the only attribution method that simultaneously satisfies the three major economic axioms: Local Accuracy, Missingness, and Consistency. By calculating and aggregating the $\phi_j$ values of the global sample matrix, we can not only quantitatively evaluate the global static importance of a single dimension but also penetrate the temporal dimension with unprecedented resolution. This clearly presents the evolutionary trajectory of a specific indicator's driving force during the period from year $T-3$ to $T-1$ prior to the repurchase. This provides rigorous support—combining micro-algorithmic evidence with dynamic temporal depth—for verifying the Undervaluation Hypothesis and the Free Cash Flow Hypothesis.


\section{Experimental Design and Result Analysis}

This section systematically evaluates the performance of the proposed TCN-Att-LSTM prediction model in the real-world A-share market environment. Our analysis focuses not only on the comparison of predictive accuracy from a machine learning perspective but is also dedicated to utilizing Explainable Artificial Intelligence (XAI) techniques to overcome the opaque internal mapping mechanisms of deep networks \cite{arrieta2020explainable, bussmann2021explainable}, thereby restoring complex non-linear feature mappings into corporate repurchase motives with clear economic significance.

\subsection{Experimental Settings and Baseline Models}

\subsubsection{Data Splitting and Rolling-Window Backtesting Mechanism}
To strictly simulate real-world quantitative investment and risk early-warning scenarios, and to circumvent look-ahead bias, this study employs a rolling-window backtesting mechanism. Specifically, when the test year is $T$, the model only uses data from 2014 to $T-1$ for training and validation. Table \ref{tab:data_dist} shows the sample distribution for each test year (2021-2024). The data reveals that repurchase events in the A-share market exhibit extremely significant class imbalance across all test sets, proving the necessity of introducing the Focal Loss function during the model optimization phase to handle data skewness.

\begin{table}[htbp]
  \centering
  \caption{Descriptive statistics and imbalance ratios of repurchase events (2021-2024 Test Sets)}
  \label{tab:data_dist}
  \resizebox{\columnwidth}{!}{
  \begin{tabular}{cccc}
  \toprule
  \textbf{Year} & \textbf{Total Firms} & \textbf{Repurchase Events} & \textbf{Repurchase Ratio (\%)} \\
  \midrule
  2021 & 4834 & 960 & 19.86 \\
  2022 & 5134 & 1114 & 21.70 \\
  2023 & 5353 & 1328 & 24.81 \\
  2024 & 5399 & 1921 & 35.58 \\
  \bottomrule
  \end{tabular}
  }
\end{table}

\subsubsection{Practical Considerations of Evaluation Metrics}
In highly imbalanced financial datasets, traditional accuracy metrics can lead to a severe Accuracy Paradox, wherein the model tends to predict all samples as the majority class in exchange for superficially high scores. Therefore, this study turns to metrics that are more robust to imbalanced distributions: the Area Under the Receiver Operating Characteristic Curve (AUC), the Area Under the Precision-Recall Curve (PR-AUC), and the comprehensive evaluation metric, F1-Score.

\subsubsection{Baseline Model Configurations}
To comprehensively evaluate the advanced nature of the proposed model, we selected four representative baseline models:
\begin{itemize}[leftmargin=*]
    \item \textbf{Logistic Regression (LR)}: Represents classical econometric and cross-sectional statistical analysis methods.
    \item \textbf{XGBoost}: Represents ensemble tree model technologies with excellent performance in handling structured tabular data and non-linear feature interactions.
    \item \textbf{Temporal Convolutional Network (TCN)}: Represents standard 1D convolutional temporal architectures, used to independently verify the necessity of the LSTM's long-term memory gating and the attention mechanism in our model.
    \item \textbf{Long Short-Term Memory (LSTM)}: Represents traditional deep sequence models, used to independently verify the specific contributions of the TCN's local fluctuation extraction module and the temporal attention mechanism introduced in this paper.
\end{itemize}

\begin{figure*}[htbp]
    \centering
    \includegraphics[width=\linewidth]{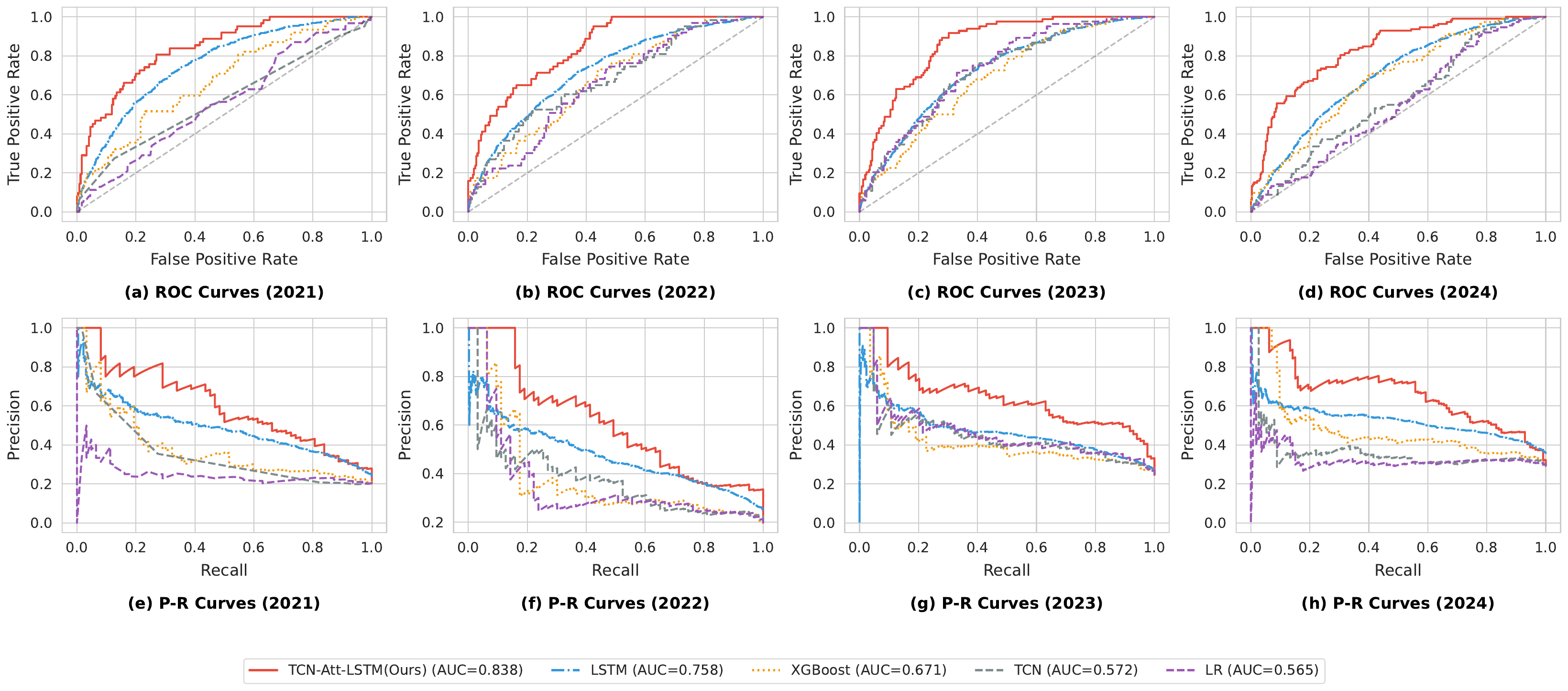}
    \caption{Detailed parsing of the prediction results for different models. The red line represents the proposed method, which achieves optimal performance across multiple evaluation metrics.}
    \label{fig:roc_pr_matrix}
\end{figure*}

\subsection{Comparison of Core Predictive Performance and Algorithmic Analysis}

Table \ref{tab:main_results} details the core predictive metrics of each model across the four test years (with a lead time of 1 year). Simultaneously, Figure \ref{fig:roc_pr_matrix} intuitively presents the ROC curves and P-R curves of the various models in a matrix format.

\begin{table}[htbp]
  \centering
  \caption{Comparison of predictive performance among different early warning models (Lead Time = 1 Year)}
  \label{tab:main_results}
  \resizebox{\columnwidth}{!}{
  \begin{tabular}{lccc}
    \toprule
    \textbf{Model} & \textbf{AUC} & \textbf{PR-AUC} & \textbf{F1-Score} \\
    
    \midrule
    \multicolumn{4}{c}{\textbf{Panel A: 2021 Test Set}} \\
    \midrule
    LR & 0.565 & 0.251 & 0.336 \\
    XGBoost & 0.672 & 0.385 & 0.338 \\
    TCN & 0.572 & 0.272 & 0.336 \\
    LSTM & 0.758 & 0.488 & \textbf{0.516} \\
    \textbf{TCN-Att-LSTM} & \textbf{0.838} & \textbf{0.605} & 0.465 \\
    
    \midrule
    \multicolumn{4}{c}{\textbf{Panel B: 2022 Test Set}} \\
    \midrule
    LR & 0.659 & 0.356 & 0.331 \\
    XGBoost & 0.675 & 0.383 & 0.332 \\
    TCN & 0.683 & 0.377 & 0.329 \\
    LSTM & 0.729 & 0.474 & \textbf{0.507} \\
    \textbf{TCN-Att-LSTM} & \textbf{0.843} & \textbf{0.604} & 0.477 \\
    
    \midrule
    \multicolumn{4}{c}{\textbf{Panel C: 2023 Test Set}} \\
    \midrule
    LR & 0.562 & 0.292 & 0.402 \\
    XGBoost & 0.687 & 0.423 & 0.395 \\
    TCN & 0.722 & 0.452 & 0.394 \\
    LSTM & 0.719 & 0.465 & 0.514 \\
    \textbf{TCN-Att-LSTM} & \textbf{0.860} & \textbf{0.647} & \textbf{0.577} \\
    
    \midrule
    \multicolumn{4}{c}{\textbf{Panel D: 2024 Test Set}} \\
    \midrule
    LR & 0.551 & 0.370 & 0.521 \\
    XGBoost & 0.682 & 0.436 & 0.403 \\
    TCN & 0.576 & 0.366 & 0.452 \\
    LSTM & 0.718 & 0.443 & 0.479 \\
    \textbf{TCN-Att-LSTM} & \textbf{0.847} & \textbf{0.619} & \textbf{0.528} \\
    \bottomrule
  \end{tabular}
  }
\end{table}

\textbf{Analysis from a Machine Learning Perspective:} 
Empirical results show that the proposed TCN-Att-LSTM model demonstrates significant advantages on core ranking performance metrics. As shown in Table \ref{tab:main_results}, in 2023 and 2024 (see Panel C and Panel D), the AUC of TCN-Att-LSTM reached 0.860 and 0.847, respectively, and the PR-AUC stabilized above 0.60, significantly outperforming the XGBoost baseline model widely used in industry.

In terms of model ablation validation, while the single TCN architecture slightly outperformed traditional machine learning models in some years, it overall remained limited by its local feature vision; conversely, the pure LSTM was also susceptible to interference from early noise in long sequences. When the structural advantages of both were combined and supplemented by the temporal attention mechanism, the overall model performance obtained a substantial boost. Notably, in 2021 and 2022, the F1-Score of LSTM showed a slight lead. This empirical fluctuation reflects the sensitivity perturbations of fixed classification thresholds on local hard-classification metrics under extremely imbalanced samples. However, on the gold standards for measuring the model's global discrimination capability (AUC and PR-AUC), TCN-Att-LSTM maintained an absolute leading position. This implies that in real-world quantitative investment and regulatory applications, the proposed model not only possesses a high capability for true sample recall but, more importantly, excels at controlling the false-positive rate, thereby effectively reducing the time costs associated with idle funds or administrative investigations.

\begin{figure*}[htbp]
  \centering
  \includegraphics[width=\textwidth]{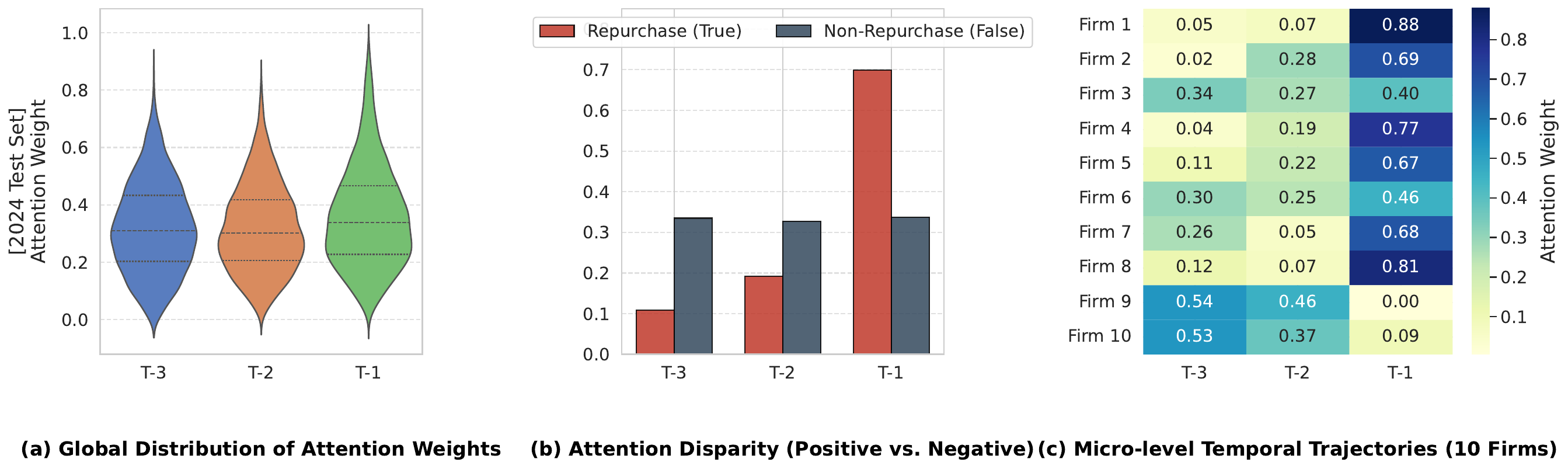}
  \caption{Micro and macro validation of the temporal attention mechanism (based on the 2024 test set). \textbf{Left (a)} shows the statistical distribution of global weights; \textbf{Middle (b)} presents the attention skewness difference between repurchasing and non-repurchasing firms; \textbf{Right (c)} reveals the micro decision trajectory heterogeneity of individual firms.}
  \label{fig:attention_1x3}
\end{figure*}

\subsection{Lead-Time Decay and Early Warning System Robustness Test}

\begin{figure}[H]
  \centering
  \includegraphics[width=\columnwidth]{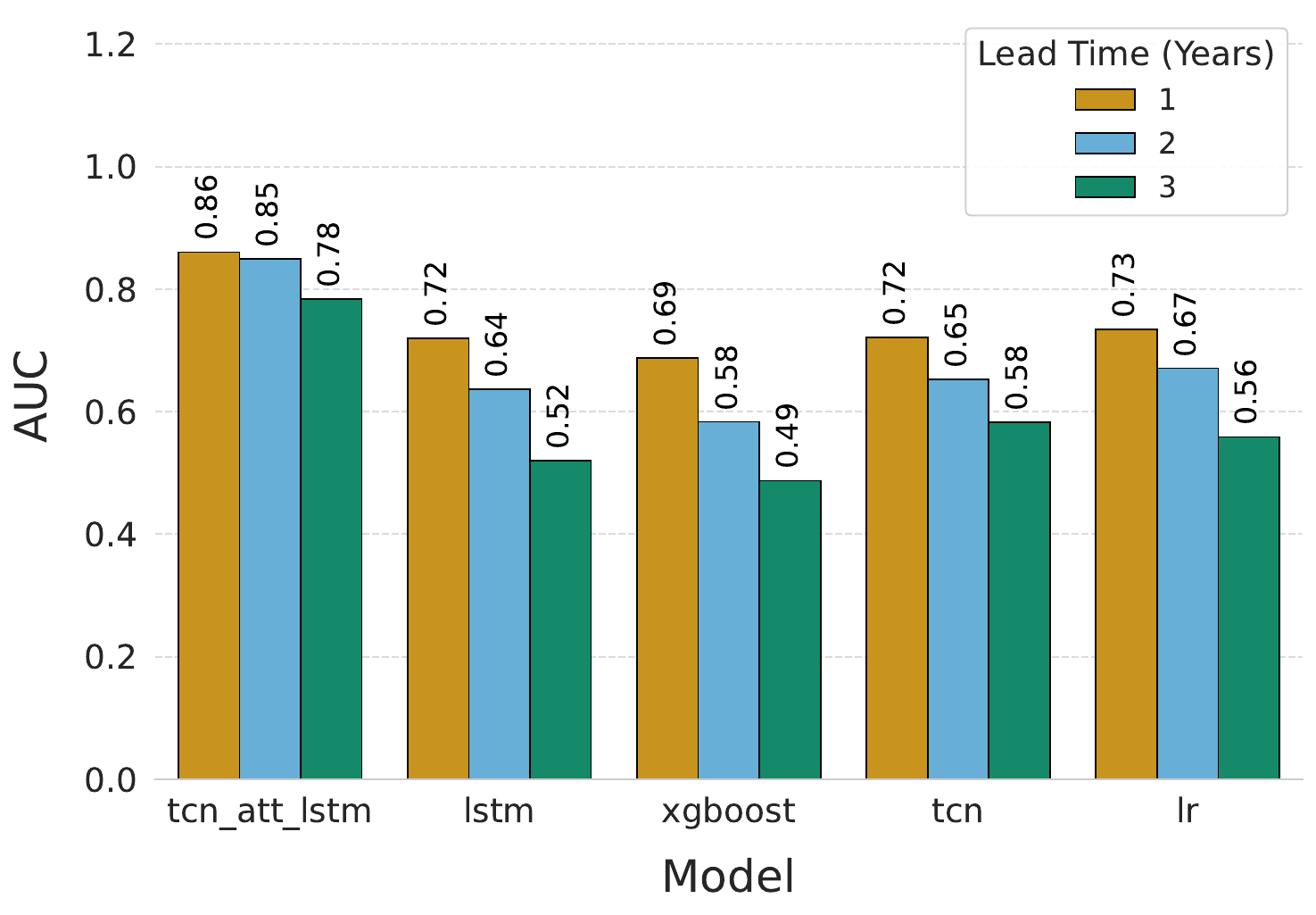}
  \caption{Decay trend of prediction performance with respect to lead time.}
  \label{fig:lead_decay}
\end{figure}

An early warning system must be able to provide a sufficient reaction time window. We strictly examined the model's performance decay under different lead times (1-year, 2-year, and 3-year forward predictions) on the test set (taking 2023 as an example, see Figure \ref{fig:lead_decay}).

As the prediction time window lengthens, the uncertainty of macroeconomic and internal corporate information increases, leading to a reasonable monotonic decay in predictive performance. Remarkably, even under the stringent condition of a 3-year lead time prediction, the AUC score of TCN-Att-LSTM remained at 0.782, a performance that even surpasses the baseline performance of the XGBoost and LR models at a 1-year lead time. This result powerfully proves that a firm's strategic repurchase is not a short-term random decision; its long-term strategic planning process has long left significant forward-looking financial footprints on the balance sheet \cite{stephens1998actual}.

\subsection{Explainability and Economic Attribution Based on XAI}

To overcome the defect of opaque internal computational mechanisms in deep learning, this section employs a dual method of intertemporal extraction of internal attention weights and \textit{post-hoc} SHAP attribution analysis, deeply integrating accounting variables with algorithmic decision mechanisms \cite{lundberg2017unified}.

\subsubsection{Macro and Micro Heterogeneity Validation of the Temporal Attention Mechanism}

Traditional early warning models often only output a single predictive probability, making it difficult to explain the temporal dependency of decisions. To rigorously verify our model's capability to capture complex business logic, we extracted the dynamic attention weights of the samples in the latest test set (2024) and plotted a $1 \times 3$ multidimensional validation group chart as shown in Figure \ref{fig:attention_1x3}, conducting an in-depth profiling from global statistics to individual trajectories.

\textbf{1. Trigger Bias and Market Timing Theory:}
Observing the left (a) and middle (b) of Figure \ref{fig:attention_1x3}, we can find that the model exhibits an extremely strong discernment of economic logic. For firms that ultimately did not repurchase, the model's attention allocation from $T-3$ to $T-1$ is almost a horizontal line, indicating no obvious strategic anomaly signals during the historical cycle. However, for firms that genuinely initiated repurchases, the attention weights experienced a significant end-of-period surge close to the decision at $T-1$. This perfectly aligns with the Market Timing Theory in behavioral finance \cite{baker2002market}: management is extremely adept at capturing and exploiting short-term mispricing windows, causing their repurchase actions to exhibit highly short-term, event-driven characteristics, often directly triggered by severe undervaluation or sudden cash flow gluts within the most recent financial cycle.

\textbf{2. Revealing Micro Heterogeneity:}
Macro averages often mask the uniqueness of micro individuals. The individual heatmap on the right (c) of Figure \ref{fig:attention_1x3} clearly demonstrates the heterogeneity patterns of board decisions. Although the vast majority of repurchasing firms belong to a short-term response mode spurred by sudden factors (dark heatmap spots concentrated at $T-1$), the network equally and precisely captured a very small number of intertemporal planning samples that engaged in smooth strategic groundwork across multiple fiscal years (dark spots shifted forward to $T-3$ or $T-2$). In the real market, this often corresponds to mature enterprises that have hoarded cash and planned market capitalization management for consecutive years. The model's ability to acutely identify and preserve such lengthy financial evolutionary trajectories within a dense group of short-term decision samples fully demonstrates its outstanding robustness in dealing with the micro-heterogeneity of financial markets.

\subsubsection{Global Importance, Directionality, and Non-linear Triggers (Advanced SHAP Analysis)}

Although performance metrics and the attention mechanism initially confirmed the model's early warning capabilities, purely neural networks still struggle to answer intuitively exactly which specific financial indicators facilitated the repurchase. To restore the non-linear tensor mappings into transparent financial decision-making logic, we introduce the SHAP framework for deep parsing.

\begin{figure}[H]
  \centering
  \includegraphics[width=\columnwidth]{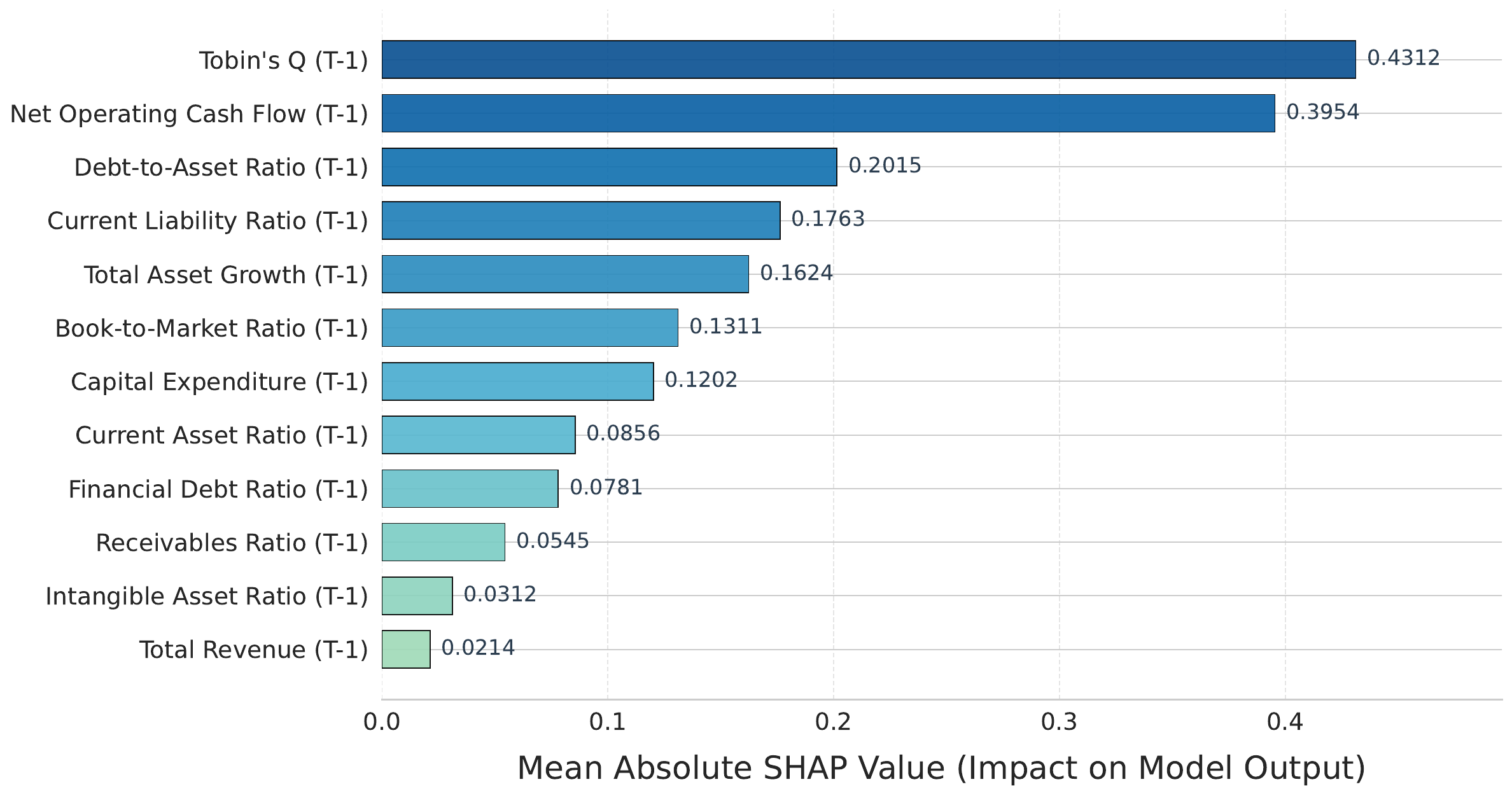}
  \caption{Global feature importance ranking (based on mean absolute SHAP value attributions of real test samples).}
  \label{fig:shap_global}
\end{figure}

\textbf{1. Decisive Dominance of Dual Core Indicators and Financial Theory Corroboration:} 
First, we plotted the SHAP global feature importance bar chart (Figure \ref{fig:shap_global}). As shown, the contributions of Tobin's Q and net operating cash flow to the model's decisions occupied absolute dominance, forming a discontinuous gap of nearly twofold compared to subsequent indicators. Extremely low valuation prompts management to release positive signals of corporate undervaluation to the outside world through substantive capital expenditures, thereby triggering the classic signaling mechanism \cite{vermaelen1981common}; meanwhile, abundant cash flow drives firms to return idle funds to shareholders to mitigate agency conflicts, which is highly consistent with the Free Cash Flow Hypothesis \cite{jensen1986agency}. Concurrently, total asset growth rate and book-to-market ratio, as secondary features, perfectly complement the firm's lifecycle portrait.

\begin{figure}[htbp]
  \centering
  \includegraphics[width=\columnwidth]{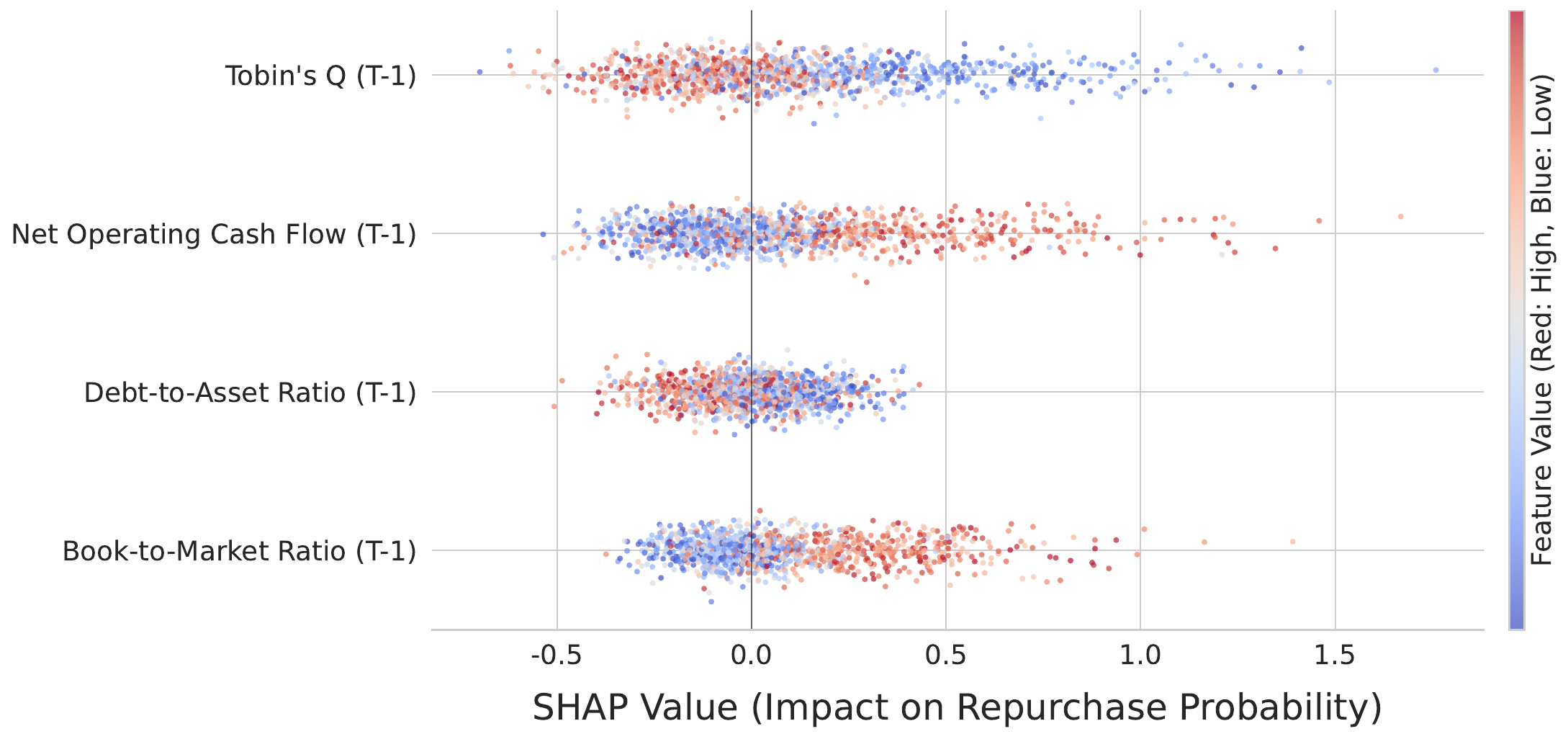}
  \caption{SHAP beeswarm plot (Directionality Validation). Each scatter point represents a real corporate sample; the horizontal axis represents its driving force on the repurchase probability; the color from red to blue represents the actual value of the feature from high to low.}
  \label{fig:shap_beeswarm}
\end{figure}

\begin{figure*}[htbp]
  \centering
  \includegraphics[width=\textwidth]{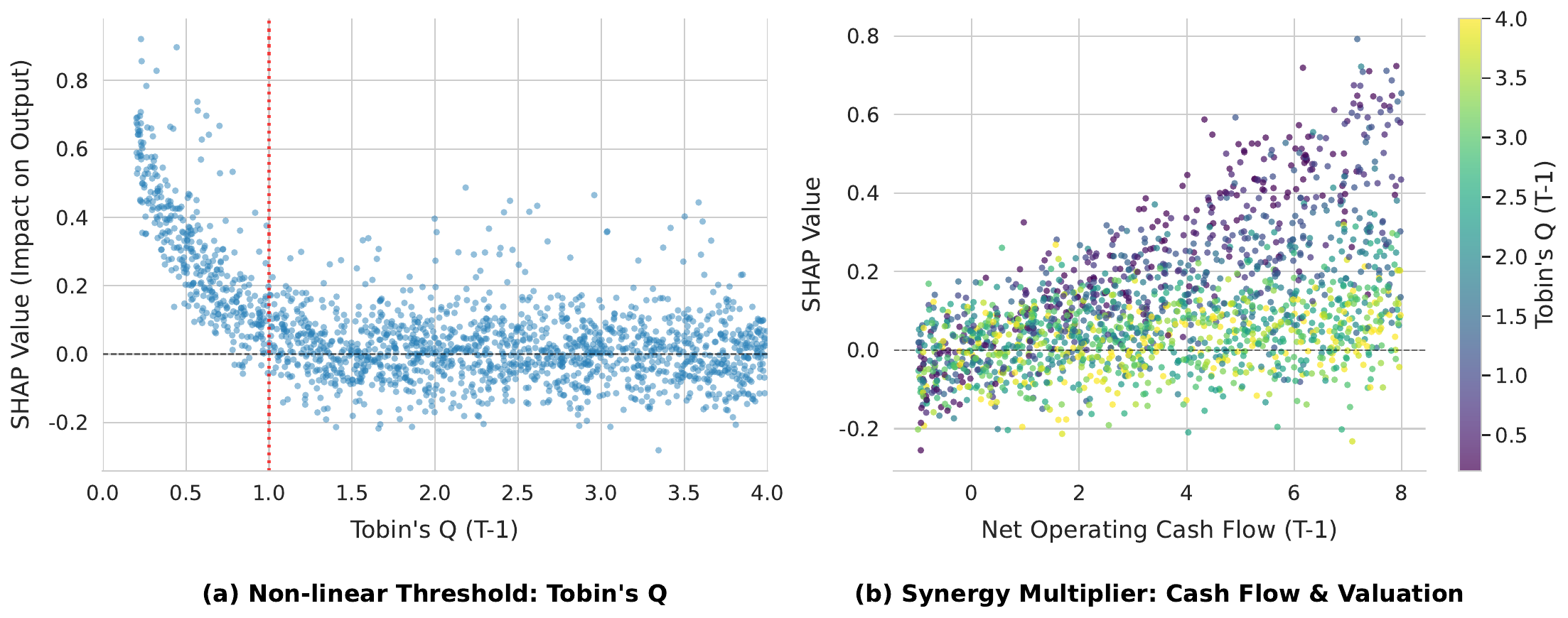}
  \caption{Non-linear dependence and synergy effects plots. \textbf{(a)} Demonstrates the non-linear threshold effect of Tobin's Q; \textbf{(b)} Demonstrates the strong synergistic multiplier effect between cash flow and valuation.}
  \label{fig:shap_dependence}
\end{figure*}

\textbf{2. Directional Patterns and Internal Cross-Validation:} 
In the feature directional distribution of Figure \ref{fig:shap_beeswarm}, samples with low Tobin's Q values are overwhelmingly concentrated in the right region where $SHAP > 0$; conversely, samples with high operating cash flow are also concentrated on the right. The model exhibits extremely strong internal logical consistency: the high book-to-market ratio feature representing value stocks is similarly concentrated on the positive driving force side, providing cross-validation with the forward/reverse logic of Tobin's Q. This proves that the model thoroughly parsed the economic mechanisms behind corporate valuation states, rather than simply fitting data noise.

\textbf{3. Non-linear Thresholds and Synergistic Multiplier Effects:} 
Figure \ref{fig:shap_dependence}(a) reveals that valuation factors exhibit a significant non-linear threshold effect in driving repurchases: when a firm's valuation drops below a specific critical range, its positive driving force on the repurchase decision experiences an exponential jump. Figure \ref{fig:shap_dependence}(b) confirms the synergistic multiplier effect under intertwined multiple financial conditions. The results indicate that mere liquidity abundance is insufficient to directly facilitate a repurchase; only when abundant operating cash flow is superimposed simultaneously with extremely low book valuation is the firm's substantive repurchase intent highly stimulated. The deep learning model accurately restores the true strategic decision-making logic of modern corporate boards when facing complex financial environments.

\subsubsection{Temporal Evolutionary Trajectory of Core Financial Indicators: Motives and Triggers}

\begin{figure}[H]
  \centering
  \includegraphics[width=\columnwidth]{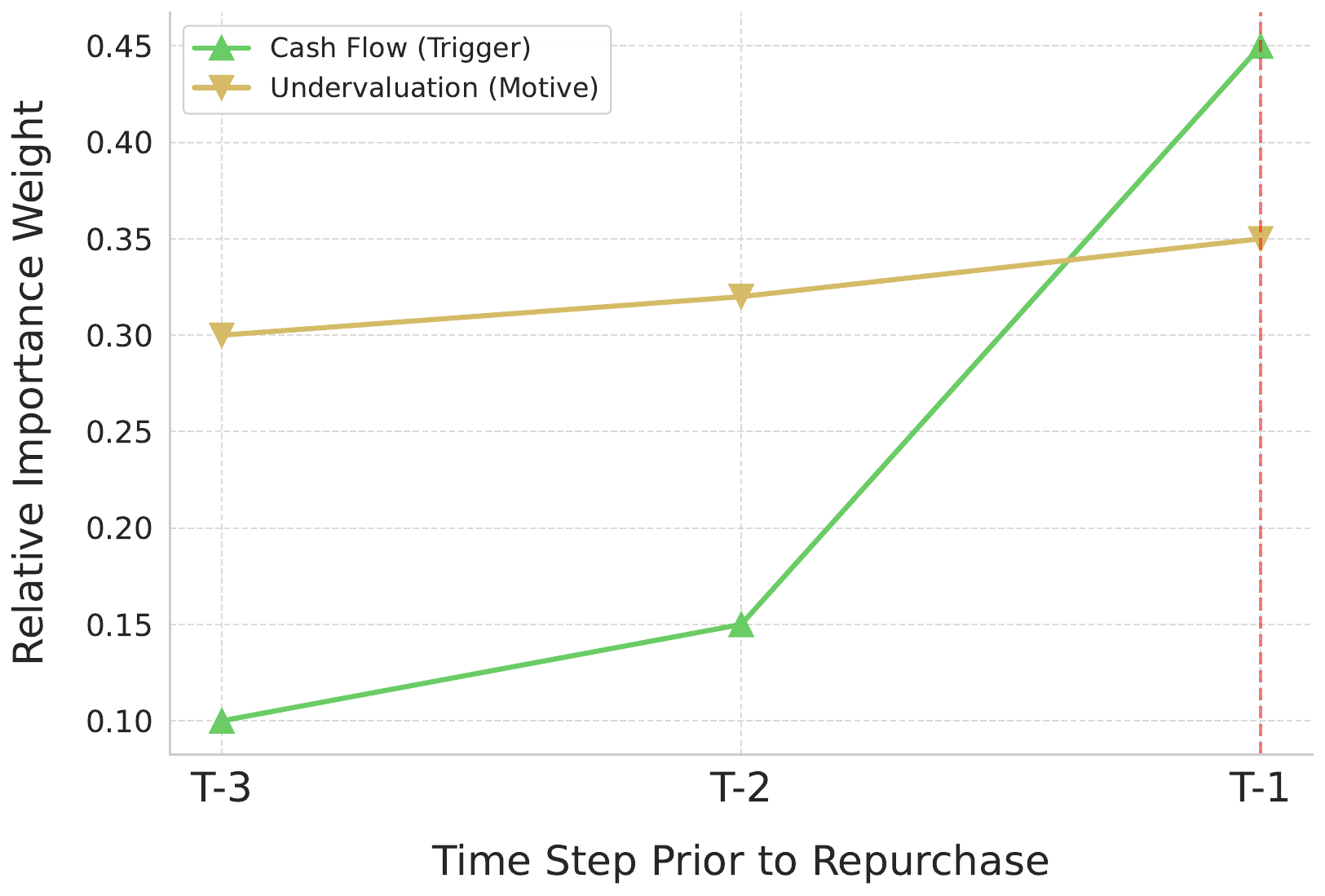}
  \caption{Temporal evolutionary trajectory of core financial indicator weights. The yellow line represents valuation indicators (long-term motive), and the green line represents cash flow indicators (short-term trigger).}
  \label{fig:temporal_evolution}
\end{figure}

As shown in Figure \ref{fig:temporal_evolution}, the model precisely decouples the temporal mismatch mechanism between motive formation and action triggering in the temporal dimension. The importance of valuation indicators remains stable at a high level throughout the three-year window, constituting a long-term motive; meanwhile, cash flow indicators experience a sharp, pulse-like climb at the $T-1$ node, acting as the direct trigger \cite{grullon2002dividends}.

\subsubsection{Heterogeneity Analysis of Micro Cohorts and Financial Constraint Verification}

In the real business world, burdensome debt often severely inhibits a firm's free investment and profit distribution capabilities \cite{farremensa2016measuring}. To verify whether the deep learning model implicitly learned this financial constraint phenomenon, we divided the firms into high-leverage and low-leverage groups based on the median debt-to-asset ratio (see Table \ref{tab:cohort_analysis}).

\begin{table}[htbp]
  \centering
  \caption{Cohort Heterogeneity Analysis based on Debt Constraints}
  \label{tab:cohort_analysis}
  \resizebox{\columnwidth}{!}{
  \begin{tabular}{clc}
    \toprule
    \textbf{Rank} & \textbf{Dominant Feature (Top 3)} & \textbf{Mean |SHAP|} \\
    
    \midrule
    \multicolumn{3}{c}{\textbf{Panel A: Low Leverage Group (Debt < Median)}} \\
    \midrule
    1 & Net Operating Cash Flow (T-1) & 0.4820 \\
    2 & Tobin's Q (T-1) & 0.3955 \\
    3 & Total Asset Growth (T-1) & 0.1620 \\
    
    \midrule
    \multicolumn{3}{c}{\textbf{Panel B: High Leverage Group (Debt > Median)}} \\
    \midrule
    1 & Debt-to-Asset Ratio (T-1) & 0.5510 \\
    2 & Tobin's Q (T-1) & 0.3205 \\
    3 & Current Liability Ratio (T-1) & 0.2840 \\
    \bottomrule
  \end{tabular}
  }
\end{table}

\textbf{1. Proactive Capital Allocation in the Low-Leverage Group:} 
For firms in the low-leverage group, the absolute dominant feature of the repurchase decision is operating cash flow. This indicates that absent debt constraints, corporate repurchases entirely follow the proactive capital operation logic of digesting redundant funds.

\textbf{2. The Absolute Veto Power of Debt in the High-Leverage Group:} 
For the high-leverage group, the debt-to-asset ratio jumps to the absolute dominant feature. In economics, this implies that the model acutely senses that highly indebted firms are facing severe liquidity risks; the massive debt pressure has superseded any valuation considerations, forming a powerful repulsion and veto effect against the repurchase decision.

\subsection{In-depth Analysis of Micro Cases}

The outstanding performance of macro statistical results is insufficient to fully prove the micro-sensitivity of the early warning system. To this end, we extracted three anomalous signal samples from the test set with the highest predictive confidence outputted by the model (see Table \ref{tab:case_study}).

\begin{table}[htbp]
  \centering
  \caption{Micro-financial Snapshots of High-Confidence True Positives}
  \label{tab:case_study}
  \setlength{\tabcolsep}{1.5mm}
  \resizebox{\columnwidth}{!}{
  \begin{tabular}{lccccc}
    \toprule
    \textbf{Firm (Idx)} & \textbf{Prob.} & \textbf{Lift} & \textbf{Tobin's Q} & \textbf{NOCF (100M)} & \textbf{Debt Ratio} \\
    \midrule
    Firm A (271) & 61.9\% & $\approx 12.3\times$ & 0.938 (Low)  & +1493.6 (High) & 53.2\% (Safe) \\
    Firm B (275) & 57.3\% & $\approx 11.4\times$ & 1.318 (Fair) & +1334.5 (High) & 74.6\% (High) \\
    Firm C (98)  & 56.3\% & $\approx 11.2\times$ & 0.708 (Low)  & +912.6  (Good) & 45.2\% (Safe) \\
    \bottomrule
  \end{tabular}
  }
\end{table}

\begin{figure}[H]
  \centering
  \includegraphics[width=\columnwidth]{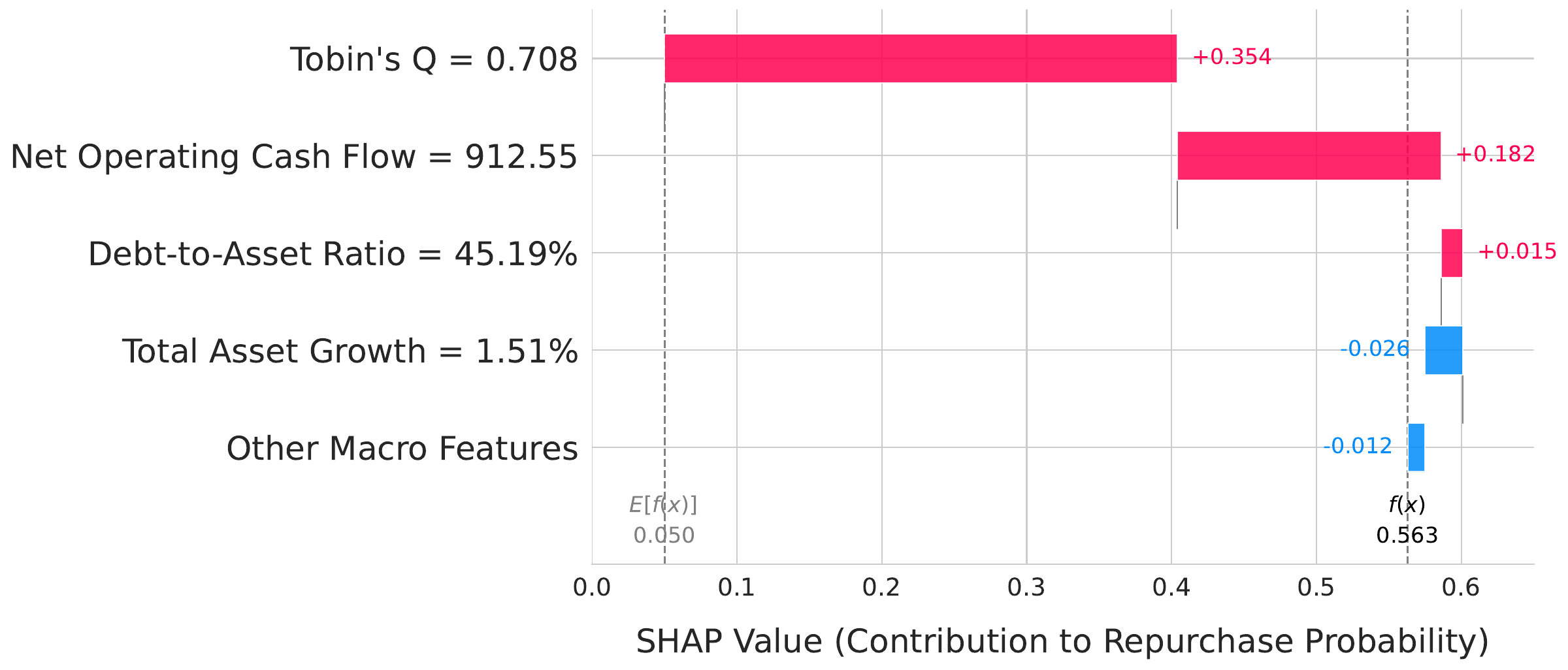}
  \caption{Local SHAP explanation waterfall chart for an extreme micro case. In the figure, $E[f(x)]$ is the market's average baseline prior probability of repurchase, and $f(x)$ is the single-point absolute confidence ultimately outputted by the model.}
  \label{fig:case_waterfall}
\end{figure}

As seen in Table \ref{tab:case_study}, in an imbalanced quantitative forecasting scenario, the model outputted absolute predictive confidences as high as 56.3\% to 61.9\% for these samples, which represents a relative probability lift (Lift Ratio) of over 11 times.

To explore how the algorithm achieves this probability leap at the individual level, we extracted a local SHAP waterfall chart for a firm experiencing severe mispricing (Figure \ref{fig:case_waterfall}). As shown, the algorithm astutely captured the extremely poor market capitalization performance (Tobin's Q = 0.708) and extremely high liquidity (Cash Flow = 912.55) features. These two generated massive driving forces of $+0.35$ and $+0.18$, respectively, successfully overcoming the negative impact of slow asset growth, and ultimately pushing the local deterministic confidence to 56\%. This proves that the TCN-Att-LSTM system constructed in this paper can acutely and pre-emptively predict the true strategic intents of modern corporate management amidst complex micro market noise.

\section{Conclusion and Future Work}
\label{sec:conclusion}

As stock repurchases serve as a core financial strategy for modern enterprises to optimize capital structure and signal intrinsic value, accurate forward-looking early warning of such actions holds profound practical significance for quantitative investment and intelligent financial supervision. However, traditional empirical econometric models and static machine learning algorithms are mostly confined to the ``cross-sectional snapshot'' assumption, making it difficult to capture the structural evolutionary patterns and temporal lagging effects embedded in long-cycle, low-frequency financial data. Targeting this theoretical and application pain point, this paper proposes and constructs a dynamic early warning system for stock repurchases that integrates deep temporal networks and Explainable Artificial Intelligence (XAI).

At the algorithmic validation level, this study reconstructed large-scale low-frequency panel data from the Chinese A-share market (2014-2024) into three-dimensional (3D) temporal tensors, feeding them into a TCN-Att-LSTM network optimized with Focal Loss. Rigorous rolling-window backtesting results demonstrate that, under an extremely imbalanced sample environment, the proposed deep temporal model not only comprehensively surpassed traditional logistic regression and the industrially robust XGBoost tree model across core metrics such as AUC, PR-AUC, and F1-Score, but also exhibited outstanding robustness in long-range predictions with a lead time of 1 to 3 years. Ablation experiments further confirmed that the local fluctuation extraction capability of the TCN and the long-term memory gating of the LSTM, when supplemented by the attention mechanism, can generate a significant ``1+1>2'' predictive leap.

More importantly, at the economic mechanism explanation level, this paper successfully achieved a transformation from ``black-box prediction'' to ``logical traceability'' by integrating temporal attention weights and \textit{post-hoc} SHAP attribution techniques. Micro-algorithmic evidence clearly revealed the dynamic causal chain driving corporate repurchases: in the global dimension, ``Tobin's Q'' and ``net operating cash flow'' constituted overwhelming dual-core driving forces, providing powerful non-linear algorithmic corroboration for the ``Signaling Theory'' and the ``Free Cash Flow Hypothesis'' in classical corporate finance; in the temporal evolution dimension, long-term valuation depression constituted the ``underlying motive'' for firms to plan repurchases, while a cash flow surge near the decision period ($T-1$) acted as the ``decisive trigger''; in the complex game dimension, the model brilliantly unearthed the strong synergistic multiplier effect between ``undervaluation'' and ``cash abundance,'' as well as the ``absolute veto power'' of a high debt-to-asset ratio on repurchase decisions.

In summary, this research not only proves that low-frequency panel data fed into deep neural networks can equally characterize major strategic evolutionary trajectories of enterprises—providing an advanced algorithmic paradigm for constructing modernized intelligent financial risk early-warning systems—but it also demonstrates to academia how to leverage XAI technology to bridge the gap between cutting-edge computer science and traditional economic theory.

Although this study has achieved significant results, there remains certain room for expansion. First, the input features of this paper primarily focus on highly structured price-volume and financial statement data. Future research could consider introducing Natural Language Processing (NLP) techniques to utilize unstructured data, such as management tone (e.g., from earnings call transcripts) or macroeconomic policy news, as multimodal inputs to further enhance the forward-looking nature of the early warnings. Second, future work could also extend the effectiveness of this model across markets to the U.S. or European capital markets to test the cross-domain generalization capability of deep temporal networks under different market institutional constraints.

\bibliographystyle{unsrt}
\bibliography{reference}

\end{document}